\documentclass[a4paper,10pt]{article}
\pdfoutput=1
\usepackage[utf8]{inputenc}
\usepackage{amsmath,graphics,caption,color,subcaption,cite}
\usepackage{jcappub}

\title{How Late can the Dark Matter form in our universe?}
\author[1,2]{Abir Sarkar,}
\author[3]{Subinoy Das,}
\author[1]{Shiv.K.Sethi}

\affiliation[1]{Raman Research Institute \\ Bangalore, India}

\affiliation[2]{Indian Institute of Science \\ Bangalore,India}
\affiliation[3]{Indian Institute of Astrophysics \\ Bangalore,India}

\emailAdd{abir@rri.res.in}
\emailAdd{abir@physics.iisc.ernet.in}
\emailAdd{subinoy@iiap.res.in}
\emailAdd{sethi@rri.res.in}

\abstract{We put constraints on the epoch of dark matter formation for 
a class of  non-WIMP (Weakly Interacting Massive Particle) 
 dark matter candidates. These models  allow 
   a fraction of Cold Dark Matter (CDM) to be  formed   between the epoch of   Big Bang Nucleosynthesis (BBN)  and the matter radiation equality. We show that for such models
the  matter power spectra  might get  strong suppression even on   scales 
 that could be  probed by linear perturbation theory at low redshifts. 
 Unlike the case of Warm Dark Matter (WDM), where the  mass of the dark matter particle controls the suppression scale, in  
Late Forming Dark Matter (LFDM) scenario, it is the redshift of the  dark matter formation which determines the form of  the matter power spectra.  We use  the Sloan Digital Sky Survey (SDSS)  galaxy clustering 
data  and the linear 
matter power spectrum  reconstructed from the  Lyman-$\alpha$ data   to  find the 
latest epoch of the  dark matter formation in our universe.  If all the 
observed dark matter is late forming,  we find   lower bounds on the redshift of dark matter formation $z_f > 1.08 \times 10^5 $ at 99.73 $\%$ C.L from the SDSS data and $z_f 
> 9 \times 10^5$, at the same C.L, from the Lyman-$\alpha$ data. If only
a fraction of the dark matter is late forming then we find tentative evidence of 
the presence of LFDM from the Lyman-$\alpha$ data. Upcoming data from 
SDSS-III/BOSS (Baryon Oscillation Spectroscopic Survey) will allow us to explore this issue in more detail.}

\begin{document}
  
\maketitle
 
\section{Introduction}
In spite of extensive search, the particle nature of the  dark matter (DM) is still a mystery. Until  now,  all the evidence for the dark matter has been 
obtained  purely through its gravitational effects. Many observations such as  
Cosmic Microwave Background (CMBR) anisotropies  \cite{Ade:2013zuv,Hinshaw:2012aka,Sievers:2013ica}, cosmological  weak gravitational lensing \cite{Bartelmann:1999yn}, galaxy rotation curves\cite{Begeman:1991iy}, 
x-ray \cite{Benson:2011uta} and large scale structure survey (e.g. \cite{Tegmark:2006az,Tegmark:2003ud}), etc.,  which span different length scales and  epochs of the 
universe, have all confirmed the presence of dark matter but do not throw
much light on its  fundamental nature. That is why, most of the search for the particle constituent of 
the dark matter have been driven by aesthetic reason. One such dark matter candidate is Weakly Interacting Massive Particles (WIMP). The ``WIMP  miracle'', which  has driven most of the direct and indirect
experimental searches, relies 
on the coincidence between weak scale cross-section and the dark matter freeze-out cross-section needed to produce correct relic density. Unfortunately, all the
direct \cite{Angloher:2011uu,Aprile:2010um,Aprile:2012nq,Ahmed:2010wy,Akerib:2013tjd}, indirect\cite{Adriani:2008zr,Adriani:2010rc,Adriani:2008zq,Adriani:2011xv,FermiLAT:2011ab,Ackermann:2010ij,Abdo:2009zk,Barwick:1997ig,Aguilar:2007yf} 
and collider \cite{Goodman:2010yf,Fox:2011pm} searches for electroweak WIMP  have not only produced null results but also different search results are in conflict with each other \cite{Hooper:2013cwa}. These anomalies definitely point 
 to much richer physics in dark matter sector  and  might direct us to think out of the box---in other words to look for different physics and energy scale of dark matter beyond weak scale 
and super-symmetric candidates.

 Once one encompasses non-WIMP candidates, the mass window for the dark matter opens up by many orders of magnitudes. 
 If dark matter is produced through thermal processes which is the case for most of the fermionic dark matter models, it can be as heavy as \rm{TeV} (super-WIMP \cite{Feng:2003xh}) or as low as \rm{keV} 
 (WDM) \cite{Dolgov:2000ew,Das:2010ts,Viel:2013fqw}. The lower bound arises from the so called Tremaine-Gunn bound \cite{Tremaine:1979we} which arises from the conserved phase space density of the dark matter particle 
 and its comparison with the densest packing of dark matter in the dark matter rich dwarf-spheroidal galaxy. But a scalar particle can also behave as dark matter. Either it can be 
 a heavy scalar boson of \rm{GeV} mass \cite{MarchRussell:2008yu,Patt:2006fw} or it can be ultra-light axion or axion-like particle \cite{Dias:2014osa,Jeong:2013oza} of  sub-eV mass. For the case of such low mass 
 bosons, a zero momentum condensate of scalar particle which arises due to coherent oscillation in a quadratic potential, behaves exactly as cold dark matter.
 
 In this work, we are  interested in epoch of dark matter formation rather than the mass of the DM particle.  Here we ask the question: how late  can a dark matter particle   form in our universe. Unlike the case of electroweak WIMP where dark matter formed 
 at very early epoch $T \simeq \rm{GeV}$, there are models of dark matter, where the production can take place even after the BBN. These models belong to the category of  ``Late Forming Dark Matter" (LFDM) \cite{Das:2006ht} where generally a late phase transition is associated with dark matter production. The LFDM models  are not only viable candidates of   theoretically motivated 
non-standard dark matter    they also  have the potential to solve some long standing cosmological issues with the  cold 
  dark matter (CDM). For instance, it is a  well known problem that N-body simulations based on  CDM produce  more galactic substructures (satellites) than observed \cite{Klypin:1999uc,Moore:1999nt}. Also, $\Lambda$CDM is subject
  to \textit{``too big to fail"}\cite{Garrison-Kimmel:2014vqa,BoylanKolchin:2011de} problem where the mass of  sub-halo is too large  in the milky way. 
One of the natural predictions of LFDM models is the 
 suppression of power at small scale and therefore the  LFDM scenario  can potentially alleviate these 
  issues with CDM. \\
  
  In this work we consider  a general class of  LFDM model, where a  
radiation-like component gets  converted into a  cold dark matter state  due to a late phase transition.  We compare the results of our theoretical
predictions with the  SDSS galaxy clustering data and  the linear
matter power spectrum extracted from the  Lyman-$\alpha$ data.  
 One of the most important parameter in the theory  is the 
 redshift of dark matter formation  which directly controls the suppression scale  in the linear matter 
  power spectra. The plan of the paper is as follows: 
in section 2, we  briefly review the process of  dark matter formation for a 
few well-motivated LFDM  models. In section 3, we discuss the cosmology of LFDM and  in section 4 we confront theoretical predictions with cosmological data.  We present our result in 
sec 5 and present a summary of our results and possible future 
directions  in section 6.

\section{Theory and motivation for  Late Forming Dark Matter}
In this section we discuss   how LFDM models  differ  with respect to their  production 
mechanism and the formation epoch as compared to the other dark matter candidates. 
For the case of electroweak WIMP, the dark matter is formed through freeze-out when the temperature of the universe  falls to  
$m_{\rm DM}/T \simeq 20$---so the production happened at very early times ($T \simeq \hbox{a few GeV}$) much before the   epoch of BBN. For the case of \rm{keV} sterile neutrino WDM, when it is produced through active-sterile oscillation, the production epoch is $T_{\rm pro} \simeq 150 \, \rm{MeV}$.
For the case of axion dark matter, the scalar starts its coherent oscillation when the mass of the scalar field becomes of the order of Hubble parameter $m_{\phi} \sim H(T)$. For the accepted 
mass scale of sub-eV axion $m_{a} \simeq 10^{-5} \, \rm{eV}$, the production happens at the QCD scale $T \simeq 100 \,  \rm{MeV}$ which also precedes  the   epoch of  BBN.\\ 

The  main difference between the models discussed above and 
  LFDM models  is that 
 the formation of CDM   can be as late as epoch corresponding to 
 $ T\simeq \, \rm{eV}$.
It is instructive to note that for warm dark matter (WDM) models, one 
 gets suppression in matter power owing to the free streaming effects which are 
governed by  the dark matter mass. But in our case, the suppression is controlled by the redshift of LFDM formation rather than its mass scale. This is the reason that  in LFDM models the existing cosmological 
 data directly constrains  the redshift of dark matter formation.

\subsection{Models of LFDM}

There are many models of LFDM  where the dark matter is created prior to recombination  but after the epoch of BBN. One such model is when dark matter is produced from  out of equilibrium 
decay of a long-lived charged particle \cite{Sigurdson:2003vy} prior to recombination. Before the decay the charged particle 
was coupled to baryon-photon plasma and then decays to neutral dark mater particle which is only gravitationally coupled to baryons and photons. \\ 

Though  in most of the models, linear matter power spectra gets a similar suppression at small scales, in this work,  we will focus on a specific class of late forming DM, where an excess radiation component $\Delta N_{\rm eff}$ makes a phase transition to a  dark matter state. We refer to the work \cite{Das:2006ht} for details of the dark matter production mechanism as well as procedure for getting linear matter power spectra for LFDM theories where a scalar field  starts coherent  oscillation after phase transition and behaves like CDM. There is another way for the case of fermions where a neutrino like light dark fermions can also be trapped in small dark matter nuggets (section 3 of  \cite{Das:2012kv})  and starts behaving like CDM.  In general, in these theories, one gets a higher $N_{\rm eff}$ compared to the case for standard $\Lambda$CDM cosmology. But we will also see that a tiny fractional increase in $N_{\rm eff}$ will suffice  for producing correct amount of CDM density if  the dark matter is 
formed  a few  e-folding before the  matter radiation equality. The recent constraints on $N_{\rm eff}$ from Planck and WMAP prefer the existence of a fractional  dark radiation $N_{\rm eff} = 3.62^{+0.50}_{-0.48}$ at 95 percent  C.L.  \cite{DiValentino:2013qma}. 
Thus this model is in complete agreement with  the cosmic microwave background measurements. In fact, if one starts with  a fully thermalised dark radiation-like component and  a fraction of it turns into CDM, it might leave a fractional dark radiation (equivalent to partially thermalised light \rm{eV} sterile neutrino) at the epoch of CMB which might even be preferred by data \cite{Archidiacono:2014apa}.

In this work we consider two cases: (a) scalar and (b) fermionic LFDM  for our numerical work. Both of them are   triggered by a late phase transition and  the epoch of phase transition $z_f$ remains the main parameter to be constrained in both the cases. Below we present a brief review of these two cases of LFDM. 

\subsubsection{Scalar LFDM} 
A dynamical scalar field with a potential $V(\phi)$  can be held  in a metastable minimum by thermal effects until a critical temperature is reached \cite{Das:2006ht}. After the universe cools down below the critical temperature, the scalar is released to oscillate around the minimum of a quadratic potential and starts  behaving like CDM with equation of state $w=0$.
 As discussed in \cite{Das:2006ht} it is possible to achieve it
by interactions beyond the standard model in the neutrino sector; these interactions allow the  the scalar field to be held in a  metastable minimum. Once the neutrino temperature drops below a critical value, the  LFDM is formed. One of the advantage of the LFDM appearing from neutrino dark energy theories is that the epoch of phase transition is naturally predicted to be very late and is subjected to   constraints arising from linear perturbation 
theory. The range for LFDM formation epoch arising from neutrino dark energy is given by \cite{Das:2006ht}: $1 \rm{eV}  \, \, <  T_f \, \, < 10^3 \rm{eV}$. 
The length scales corresponding to the horizon entry for  this range of 
 epochs are:  $2  \times \, 10^{-2} h Mpc^{-1} \, < k_f \, <  20 \, h \, Mpc^{-1}$.  This bound is purely theoretical assuming  natural values of the coupling constants. 

As discussed above,  even though QCD axion can not be late forming DM, 
 there are ultra light axion-like particles,
 $m \sim 10^{-20} \rm{eV} - 10^{-22}  \rm{eV}$, 
 arising from string theory that can behave as  LFDM depending on their  masses \cite{Bozek:2014uqa, Marsh:2011bf,Hlozek:2014lca}.  In these
cases, the linear matter power spectra has similar suppression as
 the LFDM,  appearing in the context of neutrino dark energy. It is interesting to note  that in recent work \cite{Wilkinson:2014ksa}  a CDM-like particle interacting with  neutrino or dark radiation can also produce a LFDM-like power spectra with damped oscillation. But in that case the dark matter is present
 from a very early epoch unlike the case we are interested in.

\subsubsection{Fermionic LFDM}

A light fermion-like \rm{eV} sterile neutrino can be trapped into a fermion nuggets by a phase transition driven by a strong scalar interaction.  Initially  the idea of fermion nugget formation was proposed in \cite{Afshordi:2005ym}. But in their work, the dark matter like nuggets form much later than matter-radiation equality. In a recent work (section 3 of \cite{Das:2012kv}), it was shown that  light sterile fermion behaving like dark radiation can be trapped  in heavy dark matter nuggets. The stability of the nugget is achieved when attractive fifth force is balanced by degenerate Fermi pressure of the light fermions inside the nuggets.  There are mainly two main equations which need to be solved to get the nugget mass, radius and density: 
\begin{eqnarray}
\phi'' + \frac{2}{r} \phi'& = & \frac{dV(\phi)}{d\phi} - \frac{d[\ln(m(\phi))]}{d\phi} T_{\mu}^{\mu} \\
\frac{dp}{d\phi} & = & \frac{d[\ln(m(\phi))]}{d\phi} ( 3 p -\rho)
\end{eqnarray}
We refer to \cite{Brouzakis:2005cj} for detailed derivation of these equation. Briefly, the first one is the Klein Gordon equation for $\phi(r)$ under the potential $V(\phi)= \lambda \phi^4$ where the fermions act as a source term for $\phi(r)$. The other equation tells us how the attractive fifth force is balanced by local Fermi pressure. The details of dark matter nugget formation with exact numerical solution and particle physics model will be reported soon in a different work \cite{das:2014}.

Once the phase transition happens,
there is a fractional drop in neutrino  degrees of freedom $N_{\rm eff}$, as the radiation component starts behaving as CDM immediately after the 
phase transition. This gives us one more parameter
of interest: $N_{\rm eff}$.

 Since the epoch of phase transition until the present, LFDM redshifts as normal CDM, one gets 
\begin{equation}
 \label{eq:y:1}
 \rho_{\rm lfdm}^{(z_f)} =  \rho_{\rm lfdm}^{(0)}(1+z_f)^3
\end{equation}
  Now assuming that  a fraction of excess radiation component  got converted into 
a fraction of CDM density, $f_{\rm lfdm}$,  at $z = z_f$, we get the decrement in the effective number of 
neutrino degrees of freedom, $\Delta N_{\rm eff}$ to be:
\begin{equation}
 \label{eq:y:2}
 \Delta N_{\rm eff} \rho_{\nu}^{(z_f)} = f_{\rm lfdm} \rho_{\rm lfdm}^{(0)}(1+z_f)^3
\end{equation}
  where $\rho_{\nu}^{z_f}$ is the energy density of one neutrino-like radiation
 species at the formation  redshift. This yields: 
 \begin{equation}
 \label{eq:y:3}
  \Delta N_{\rm eff}= f_{\rm lfdm} \frac{\rho_{\rm cdm}^{(0)}}{\rho_{\nu}^{(0)}}= 1.7 f_{\rm lfdm} \Omega_{\rm CDM} h^2 \left ({10^5 \over 1+z_f} \right )
\end{equation}
It should be noted that  $\Delta N_{\rm eff}$ is inversely proportional to the redshift
of formation.  As the effective number of neutrino degrees of freedom dynamically change  in this model,  observational constraints on $N_{\rm eff}$ from
different observations need to be interpreted properly.

For instance, for $z_f < 10^{10}$, the BBN constraints, which
depend on the in situ value of $N_{\rm eff}$ during the era of BBN, apply to
the value of $N_{\rm eff}$ before the epoch of dark matter formation.   \cite{Shvartsman:1969mm,Mangano:2011ar,2013neco.book.....L}. On the other hand, CMBR and 
galaxy clustering data, which are influenced by the history of 
changes in $N_{\rm eff}$,  are also sensitive to  the final $N_{\rm eff}$. Throughout this 
paper $N_{\rm eff}$ corresponds to the initial degrees of freedom. We also
note that, for most of the range of $z_f$ of interest, 
 $\Delta N_{\rm eff}$ is generally smaller than   the current 
precision on  $N_{\rm eff}$ from different data sets; for instance, it follows
from Eq.~\eqref{eq:y:3} that even if $f_{\rm lfdm} = 1$, $\Delta N_{\rm eff}$ = 0.2
for $z_f = 10^5$, assuming the best fit Planck parameters 
for $\Omega_{\rm CDM} h^2$. 
As noted above we also consider the case when the LFDM contributes only a fraction to 
the observed CDM at the present and  this fraction is denoted by $f_{\rm lfdm}$.

It is instructive to note that  in  both the above cases,  once $z_f$ is fixed, the power spectra is almost uniquely determined as the model has to match the correct dark matter relic density. This also means
that the resulting   constraint on  $z_f$ from the data  would be valid for both the fermionic and scalar cases of LFDM.

\begin{figure}[h!]
\centering
\includegraphics[height=5in, width=5in]{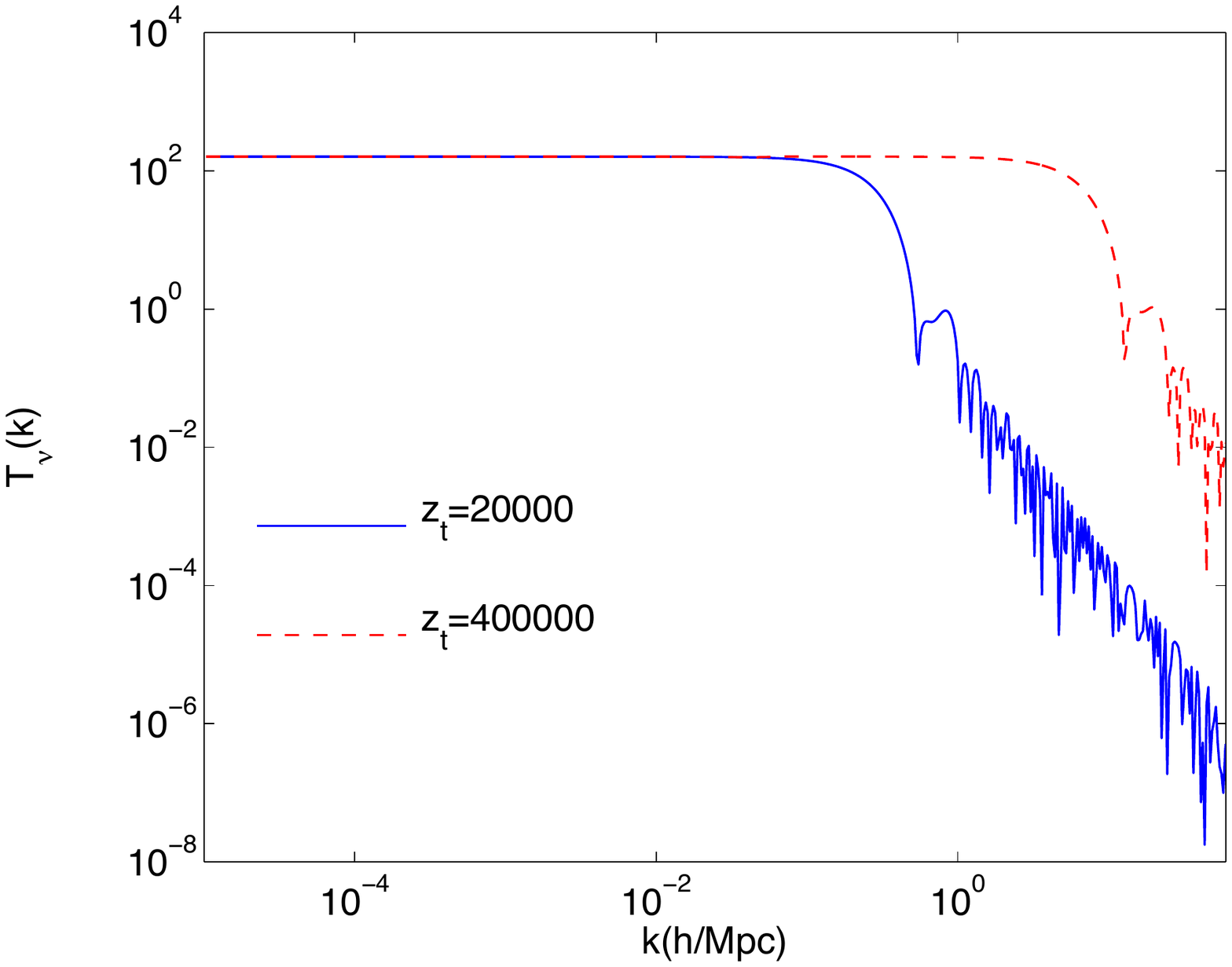}
\caption{The unnormalized transfer functions for the neutrino density perturbation $\delta_{\nu}$ at two different epochs of LFDM formation are shown. This fluctuation
 provides the initial condition for the LFDM which evolves 
 like cold dark matter from  $z_f $ until present epoch.}
\label{fig:1a}
\end{figure}

\section{Cosmology of LFDM}

There are mainly two main features of LFDM cosmology 
that manifest themselves in the  matter power spectra. First, there is a sharp break in the power  at the co-moving  scale $k = aH_e$; here $H_e$ is the Hubble scale  at the epoch of phase transition. Second, there are  damped oscillations at  smaller scales. Both of these can be seen in  Figure~\ref{fig:1a}, in which
we show the transfer function for massless neutrinos. 
Before the phase transition, due to strong coupling of the scalar with 
  neutrinos, the 
scalar  perturbations  follow neutrino perturbations. We compute the 
 the density perturbations of the  massless neutrinos at $z = z_f$  and match that to the initial conditions  of the CDM component  at that epoch. 

\subsection{Modification in CAMB}

It is  interesting to note that even though the physics of scalar and fermionic  LFDM production can be  quite different, the initial condition of the LFDM at the formation epoch can be taken from the a neutrino-like component at the epoch corresponding to $z = z_f$. This is because in both the cases, the density perturbation of a neutrino or a dark radiation component provides the initial density fluctuation of LFDM at the production epoch 
$z_f$. The evolution of neutrino density perturbation is  obtained by solving a series of coupled differential equations  \cite{Ma:1995ey} involving Legendre Polynomials
 \begin{eqnarray}
\dot{\delta} & = & -\frac{4}{3} \theta - \frac{2}{3} \dot h \nonumber \\
\dot{\theta} & = & k^2\left(\frac{\delta}{4} - \sigma \right) \nonumber \\
2 \dot{\sigma} & = & \frac{8}{15} \theta - \frac{3}{15} k F_3
+ \frac{4}{15} \dot h + \frac{8}{5} \dot{\eta} \nonumber \\
\dot{F}_l & = & \frac{k}{2l+1} \left(l F_{l-1} - (l+1) F_{l+1}
\right)
\end{eqnarray}
 The  solution for $\delta_{\nu}$ is an exponentially damped  oscillation at sub-horizon
scales \cite{Ma:1995ey}.  Physically, it represents the free-streaming effects of  highly relativistic neutrinos. In Figure~\ref{fig:1a}, we plot transfer function for standard model neutrino density fluctuation by solving the above equations using the publicly available code \texttt{CAMB}\cite{Lewis:1999bs} for two different values of  $z_f$. Our main modification  in \texttt{CAMB} is  to  evolve it up to a redshift $z_f$ without  CDM  and extract the transfer function for  neutrino perturbation  at $z = z_f$,  $ \delta_{\nu} (z_f)$,  and use that for the LFDM initial condition for density fluctuation at the epoch of its formation. We then  evolve LFDM perturbation  just like CDM to get the power spectra at the  present epoch.  So the oscillations  at  small scales in the final power spectra at $z=0$ is a signature of the fact that LFDM obtained its initial density fluctuation from neutrino perturbation at $z_f$ which was damped and oscillatory at scales smaller than the horizon size of the Universe at    $z = z_f$.

The main goal of our  work is to find out how late the  dark matter can
 form,  i.e,  to
 find out the minimum value of the formation  redshift $z_f$.  As discussed in 
the previous section, the formation of dark
matter happens via the transition of the scalar field. 
We are therefore able to formulate the cosmological impact of LFDM in terms
of three parameters: the initial relativistic 
neutrino degrees of freedom: $N_{\rm eff}$, the epoch of the formation
of CDM: $z_f$,  and the fraction of CDM that forms at $z = z_f$: $f_{\rm lfdm}$.  

\begin{figure}
\begin{subfigure}{.5\textwidth}
  \centering
  \includegraphics[width=1.0\linewidth]{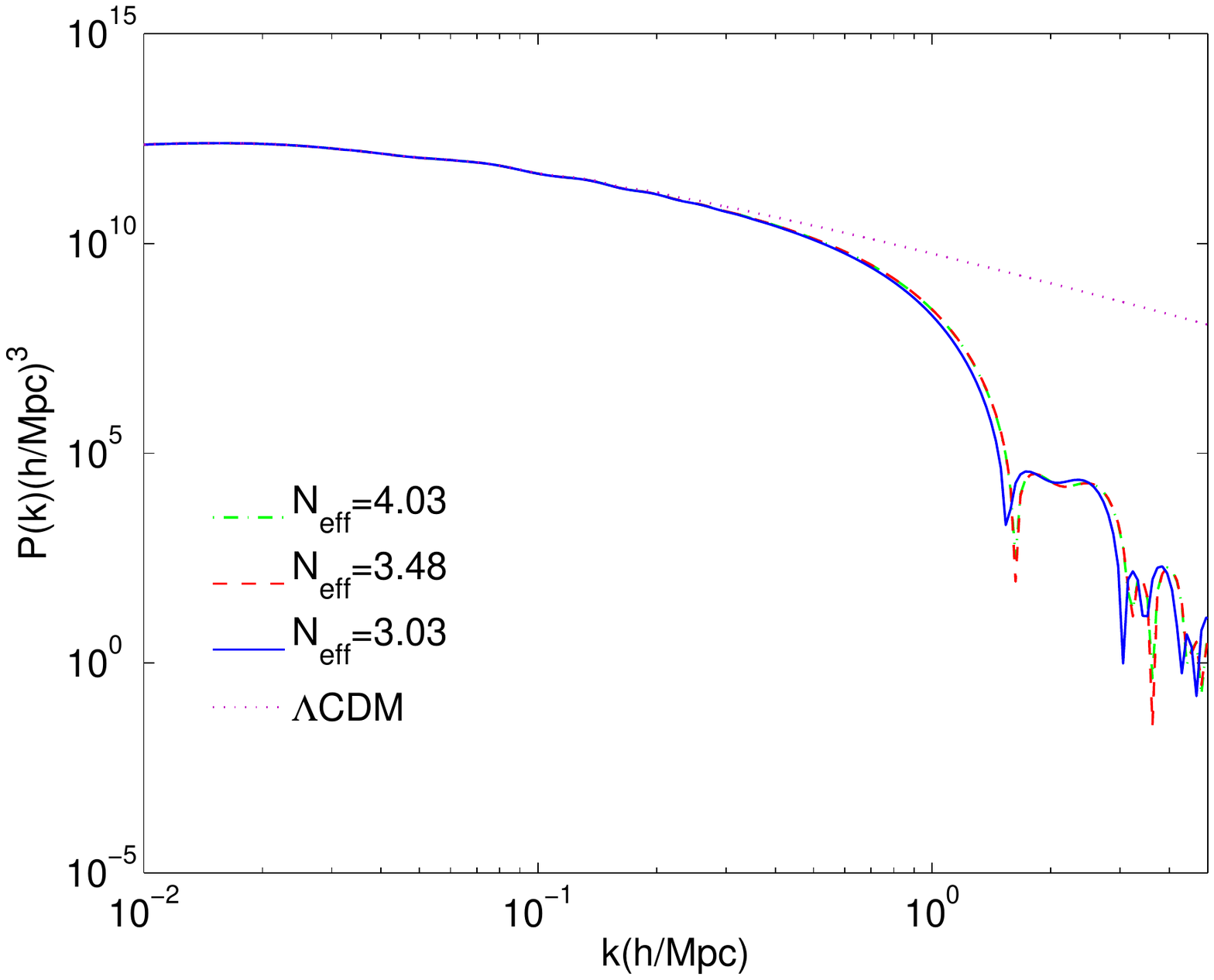}
  \label{fig:1q}
\end{subfigure}%
\begin{subfigure}{.5\textwidth}
  \centering
  \includegraphics[width=1.0\linewidth]{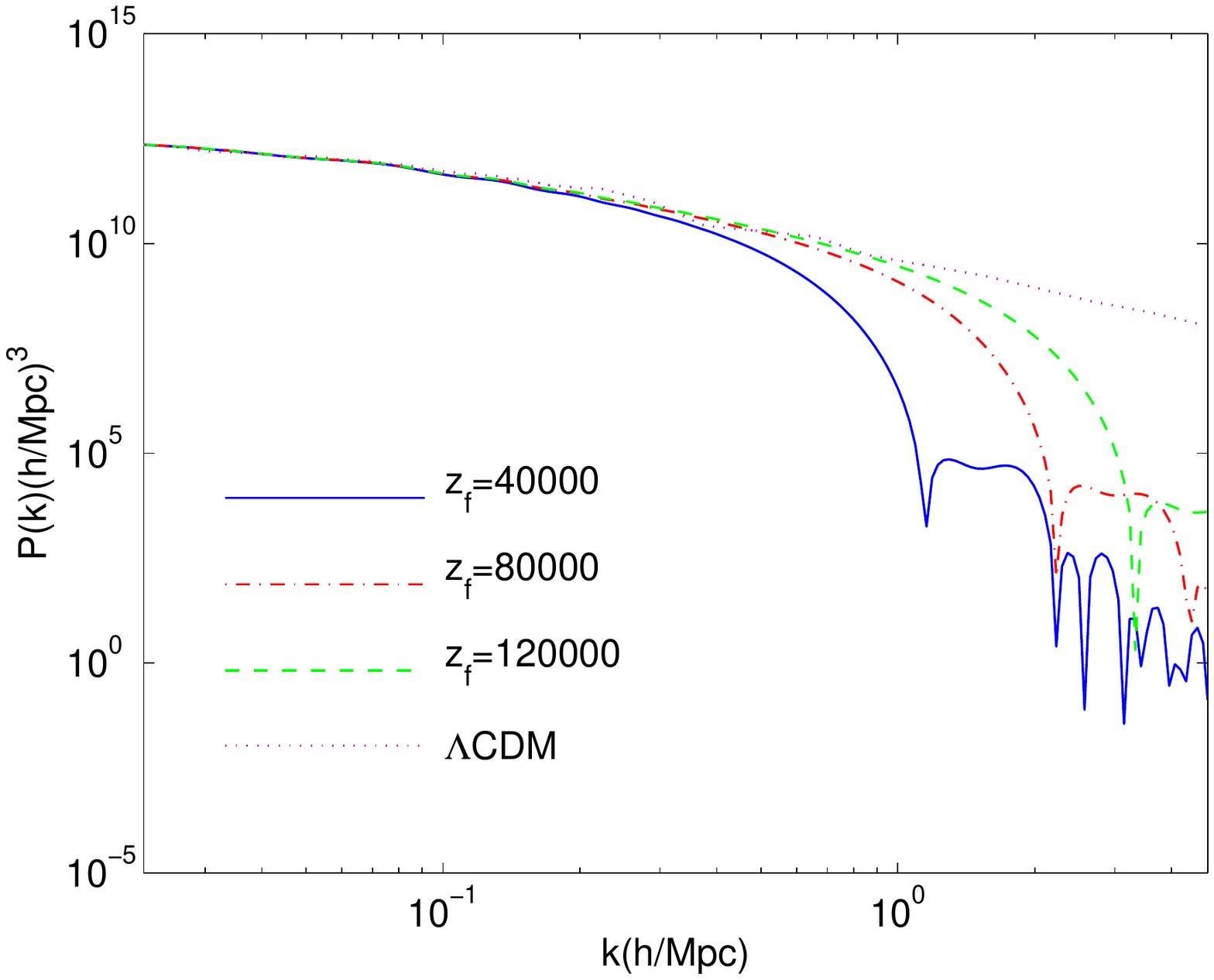}
  \label{fig:1w}
\end{subfigure}
\caption{The  LFDM power spectra (unnormalized) are shown for a range of 
formation  redshift $z_f$ and relativistic neutrino degrees of freedom. 
The left panel shows the impact of changing  $N_{\rm eff}$ for a fixed
 $z_f=52000$. In the right panel, LFDM power spectra
are shown  for different  $z_f$ for $N_{\rm eff} = 3.04$.}
\label{fig:1}
\end{figure}

A set of power spectra with different  $z_f$  and  $N_{\rm eff}$ 
are shown in Figure~\ref{fig:1}. 
In each case we have plotted the usual  $\Lambda$CDM power spectrum for comparison.

As Figure~\ref{fig:1} shows, the new features introduced by  LFDM are largely determined by 
 the variation of $z_f$. The scale
imprinted on the matter power spectrum is determined by the scale of horizon
entry at $z =z_f$, $k_e$. For the horizon entry in the radiation dominated era:
\begin{equation}
k_e = {H_0 \over c}  (1+z_f)\Omega_\gamma (1+0.227N_{\rm eff})
\end{equation}
Here $\Omega_\gamma$ is the radiation contribution from photons.  The matter power
spectrum is suppressed at scales below the corresponding scale for $k_e$. 
This suppression can be understood as follows: the LFDM obtains
its initial conditions from  massless neutrinos. On the super horizon 
scales, the massless neutrinos behave like  other forms of 
matter  such as the CDM  (for details see e.g. \cite{Ma:1995ey,Peebles:1970ag}). However, unlike CDM, the perturbations in this component are washed out
owing to free-streaming on scales smaller than horizon size. 
 As $z_f$ is increased the feature shifts to larger $k_e$, or smaller scales. 
As $z_f$ tends to infinity, the LFDM matter spectrum approaches the $\Lambda$CDM results. This also motivates our choice of the cosmological data for constraining the LFDM model.

\section{The Data}
As discussed above, we can theoretically analyze the 
impact  of the late forming dark matter
in terms of three  parameters: $z_f$, $f_{\rm lfdm}$ and $N_{\rm eff}$. For a given  $z_f$ and $f_{\rm lfdm}$,  $\Delta{N_{\rm eff}}$ can be expressed in terms of 
these parameters for a given $\Omega_{\rm DM}h^2$ (Eq.~\eqref{eq:y:3}), 
which we assume to be fixed and given by the best-fit Planck estimate
for the six-parameter spatially-flat $\Lambda$CDM model \cite{Ade:2013zuv}.

The two parameters---$z_f$ and $N_{\rm eff}$---affect the linear power spectrum at different scales. The 
main impact of changing $N_{\rm eff}$ is to alter the matter radiation equality
epoch. This shifts the peak of the matter power spectrum. 
As the SDSS data on the galaxy  power spectrum gives   the  power at such
scales: k=$0.02\hbox{--}0.1 \, \rm h/Mpc$, this data is sensitive to the variation of  of $N_{\rm eff}$. We use  the SDSS DR7 release data \cite{Reid:2009xm}. For $k > 0.1$, the SDSS 
data cannot be directly compared to the predictions of linear theory as 
non-linearities set in for such scales. We use the HALOFIT model  embedded in
 \texttt{CAMB} to obtain the  non-linear power spectrum 
for comparison with  the SDSS galaxy power spectrum;  
this procedure allows us to use the data for $k \lesssim 0.2$.  It is 
instructive to note that though the HALOFIT works mainly for $\Lambda$CDM and
 might not work  for other dark energy models of constant equation 
of state differing from $w =-1$ \cite{Smith:2002dz}, in  LFDM cosmology, the background 
 evolution is exactly same  as  the $\Lambda$CDM  model after the phase transition
 has occurred deep in radiation dominated era. So we expect HALOFIT
 to be a good approximation for mildly non-linear power spectra for comparison
 with SDSS galaxy power spectrum.

As seen in Figures~\ref{fig:1a} and~\ref{fig:1} above, the main effect of late
formation redshifts  $z_f$ is to suppress
the power at  scales  $k>0.1 \rm \, h/Mpc$. Such scales are not directly accessible 
from the data on galaxy power spectrum at low redshifts. It is known that
Lyman-$\alpha$ clouds observable at intermediate redshifts ($2 < z < 5$) 
probe mild over densities ($\delta  \simeq 10$) of  the density field. The data
from Lyman-$\alpha$ clouds can be used to reconstruct the linear matter 
power spectrum for scales comparable to the Jeans' scale of the intergalactic 
medium in the relevant redshift range \cite{Croft:2000hs,Gnedin:2001wg}.
Here we use the data in the range: $0.2 < k < 4.8 \, \rm h/Mpc$ from
 \cite{Croft:2000hs,Tegmark:2002cy}.   From Figure~\ref{fig:1a} and~\ref{fig:1}, it is clear that the  scales probed by the Lyman-$\alpha$ data 
 are  far more sensitive to the variation
in $z_f$. As $z_f$ is increased the oscillations seen in the power spectra
move to larger values of  $k$ (or smaller scales)  with the power spectrum approaching the $\Lambda$CDM model
as $z_f$ tends to very large values. 

Other data sets at  scales overlapping with SDSS data 
 are available, e.g. WiggleZ survey \cite{Parkinson:2012vd} with scale 
coverage $0.01 < k < 0.5 \, \rm h \, Mpc^{-1}$. We could obtain supplementary
information from WiggleZ data but it doesn't 
expand the range of scale we already consider. 
 Or the two data sets we use allow 
us to obtain the tightest possible constraints  on LFDM models within the 
framework of linear (and mildly non-linear) theory. Cosmological weak 
lensing provides a powerful  probe of the matter power spectrum (e.g. \cite{Heymans:2013fya}). We do not use it here because the scales probed by the cosmological  lensing are larger than the those probed by the Lyman-$\alpha$ data (e.g. \cite{Tegmark:2002cy})  so we cannot use it to get 
better constraints on the  formation redshift $z_f$. Also in this paper 
we only consider the available data on measured or reconstructed power spectra.
The reconstructed power spectrum 
 is not readily available in the literature (e.g. \cite{Heymans:2013fya}).
This means we have to compute the observables presented in the literature 
from LFDM power spectra. We shall undertake this task in future works. 

Our choice of Lyman-$\alpha$ data is also governed by the availability of 
reconstructed linear power spectra. In all the available data on
the linear matter power spectrum, the data we use provides
a  probe of the  smallest scales. It  is based on  the high 
spectral resolution QSO spectra (total of 53 QSOs including 30 observed at high
spectral resolution \cite{Croft:2000hs}). This one-dimensional  data allows reconstruction of 
the linear 3-dimensional matter spectrum. However the low-resolution 
SDSS data, which is available  for a much larger number of QSOs,  doesn't allow this 
reconstruction  (for details and 
discussion see e.g. \cite{Croft:2000hs,McDonald:2004xn}). 
This means that a comparison with    the  ongoing survey SDSS-III/BOSS,  which will finally  obtain spectra of 
160000 QSOs in the redshift range $2 < z< 7.5$ \cite{Palanque-Delabrouille:2013gaa}, will require us 
to simulate the Lyman-$\alpha$ spectra for our class of models.  We consider 
it beyond the scope of this paper and plan to undertake this study in the 
near future. We also  note that even the low spectral resolution 
Lyman-$\alpha$ could be a powerful probe of the matter power spectrum
at small scales because of two reasons: (a) the measured 1-dimensional flux
power spectrum by the Lyman-$\alpha$ data receives contribution from
a wide range of scales of the 3-dimensional power spectrum, (b) the relation
between the density field and the observable is non-linear (e.g. \cite{Pandey:2012ss}).

\begin{figure}
\begin{subfigure}{.5\textwidth}
  \includegraphics[width=1.0\linewidth]{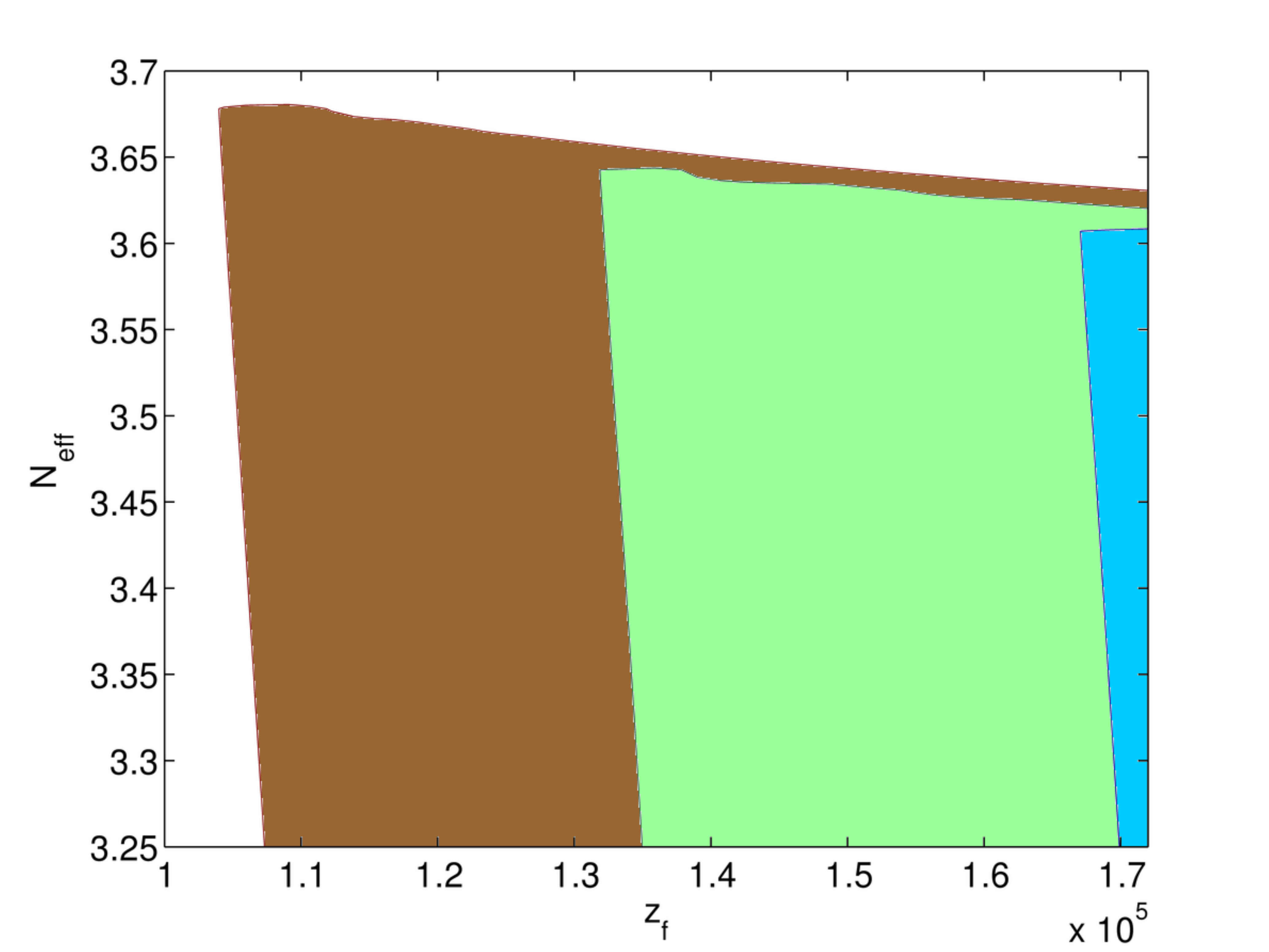}
  \label{fig:2a}
\end{subfigure}%
\begin{subfigure}{.5\textwidth}
  
  \includegraphics[width=1.0\linewidth]{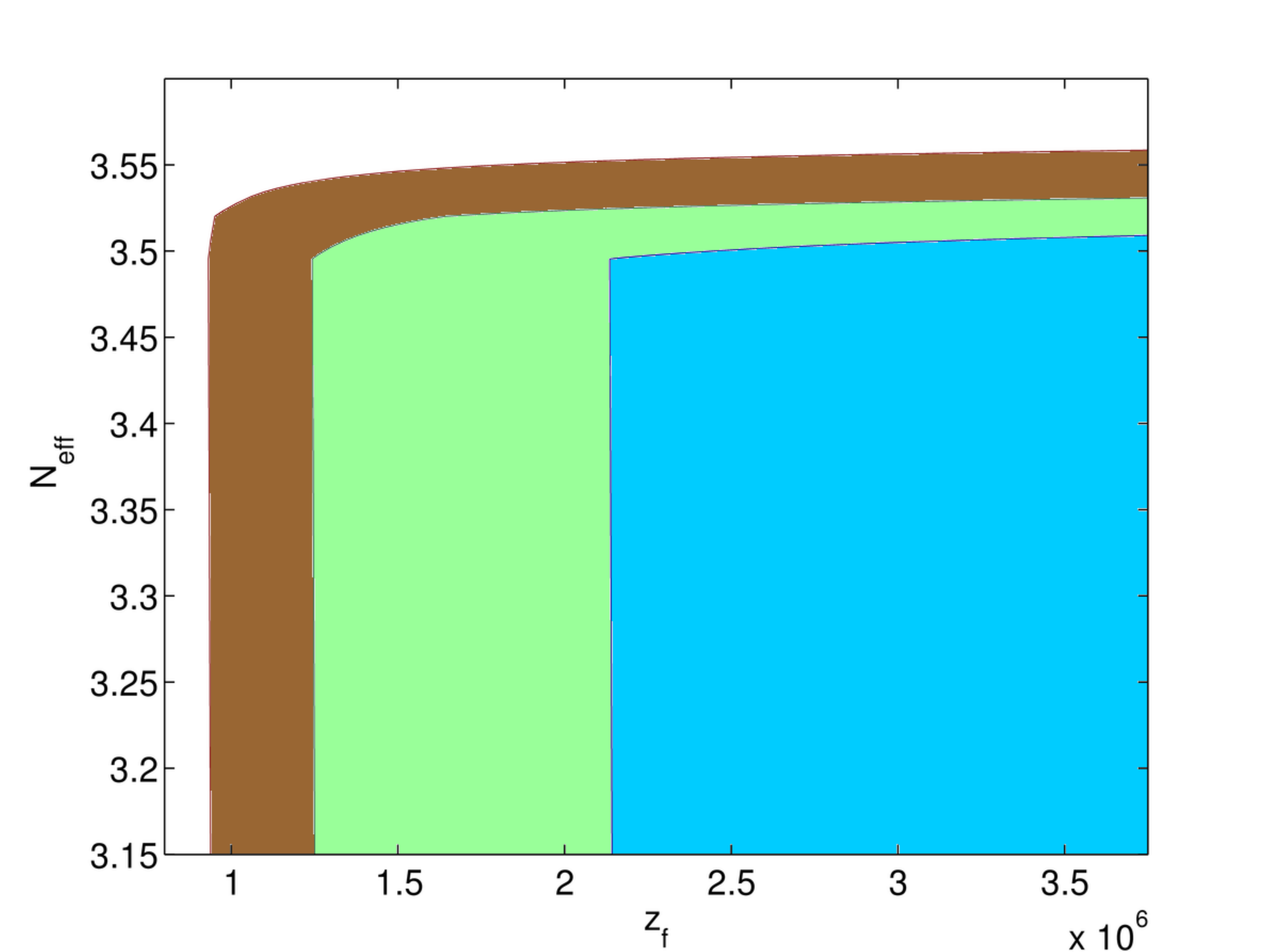}
  \label{fig:2b}
\end{subfigure}
\caption{The allowed regions in the  $z_f$--$N_{\rm eff}$ plane 
are shown from the power spectra of  SDSS (Left Panel) and Lyman-$\alpha$ (Right Panel) data. In each Panel,  the blue, green and the brown regions
 indicate the 68$\%$, 95.4$\%$ and 99.73$\%$ confidence levels, respectively.}
\label{fig:2}
\end{figure}

 \section{Data Analysis and Results}
  The two data sets---SDSS galaxy power spectrum and the linear power spectrum
reconstructed from the Lyman-$\alpha$ data---allow us 
to investigate the range of scale: $k=0.02\hbox{--}5 h/Mpc$. However, these 
two data sets do not have the same bias with respect to the underlying density
field, and therefore the overall normalization constant is different for
the two cases. In other words, we can probe the shape of the power spectrum
in the aforementioned range of scales and not its overall normalization. We consider four  parameters for each data set: $N_{\rm eff}$, $z_f$, $f_{\rm lfdm}$, 
and $C$, where $C$ corresponds to an overall normalization which is marginalized. For our analysis we search the best-fit in the range:  $N_{\rm eff} = 3\hbox{--}4$, 
which encompasses   the current range of  constraints on $N_{\rm eff}$ \cite{Mangano:2011ar,2013neco.book.....L}. 

As noted above we compute a suite of models for different $z_f$ and $N_{\rm eff}$
by modifying \texttt{CAMB}. We extract unnormalized power spectra and the normalization
 is fixed by comparison with data. Model  predictions for a range of 
$f_{\rm lfdm}$ are obtained by assigning different weights to the initial conditions; for instance, for $f_{\rm lfdm} = 1$, the initial condition for the CDM component is  drawn from massless neutrinos at $z= z_f$. For a smaller value of $f_{\rm lfdm}$, the initial conditions are a mix of the CDM component in the pre-transition
phase and the massless neutrino. This also means that we need to vary only 
two parameters  ($z_f$ and $N_{\rm eff}$) in CAMB for obtaining the power spectra
for all the four parameters. For  likelihood analysis  we have used the range of  $z_f$ to be $24000\hbox{--}180000$ with an interval of $\Delta z_f=4000$ while analyzing 
the SDSS data and $z_f= 62000\hbox{--}4000000$ with an interval of $\Delta z_f =2000$ for the Lyman-$\alpha$ data. The range of the $z_f$ is different for the 
two data-sets because the Lyman-$\alpha$ data covers much smaller scales as 
compared to the  SDSS data. The smallest scale probed 
by the Lyman-$\alpha$ data,  $k \simeq 4 \, \rm Mpc^{-1}$ enters the horizon at  $z \simeq 4000000$ which is the highest $z_f$ we have considered. Similarly, 
$N_{\rm eff}$ is also finely sampled to ensure convergence of the likelihood
procedure. 

We use 45 band-powers  from the SDSS galaxy data and 12 points from the reconstructed
linear power spectrum from the Lyman-$\alpha$ data.  The best-fit $\chi^2$
for the two case is $65$ and $10.5$, respectively.  
 The multi-parameter contours and posterior probabilities  are  computed by 
marginalization, i.e. the integration of 
the  likelihood function  $\exp (-\chi^2/2)$ over redundant parameters.

We first consider the case $f_{\rm lfdm} = 1$, or all the observed CDM at the 
present is formed at $z_f$. 
 In Figure~\ref{fig:2}, we show  the confidence limits for $z_f$ and $N_{\rm eff}$
for the two data sets. Both the data sets result in a lower limit on the 
value of $z_f$. The Lyman-$\alpha$ data results in stronger constraints on $z_f$.  This result follows from Eq.~\eqref{eq:y:3} and Figure~\ref{fig:1}
which show that  an increase in $z_f$ results in the feature in the power spectrum
shifting to smaller scales. As Lyman-$\alpha$ data  probe smaller
scales, we expect a tighter constraint on $z_f$ from these observations. 
We note that for both the data sets the floor on the value of $\chi^2$  is 
set by the $\Lambda$CDM model. Or we do not find any evidence of an improvement
over the $\Lambda$CDM model within the framework of a two-parameter LFDM model.

\begin{figure}
\begin{subfigure}{.5\textwidth}
  \includegraphics[width=1.0\linewidth]{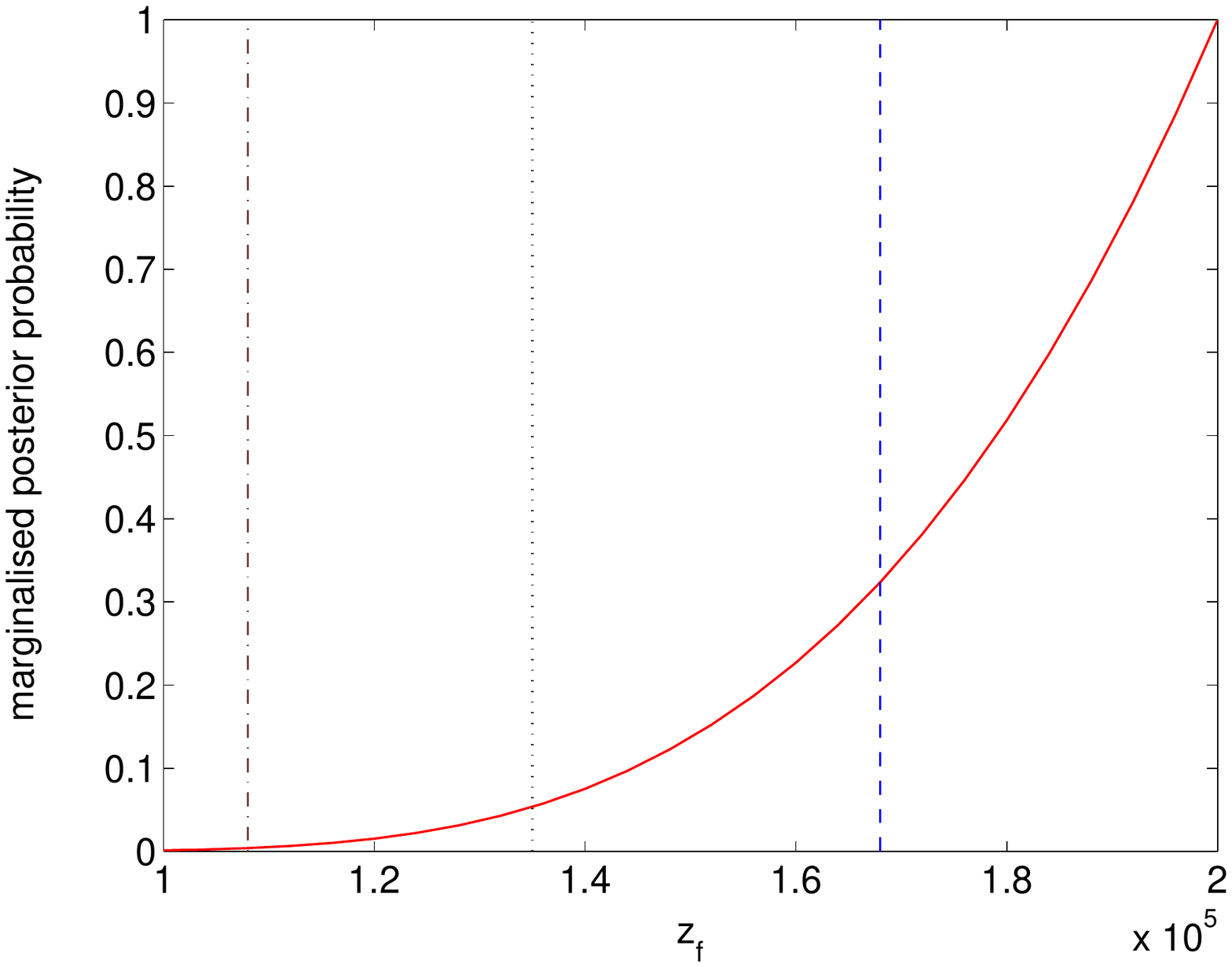}
  \label{fig:3a}
\end{subfigure}%
\begin{subfigure}{.5\textwidth}
  
  \includegraphics[width=1.0\linewidth]{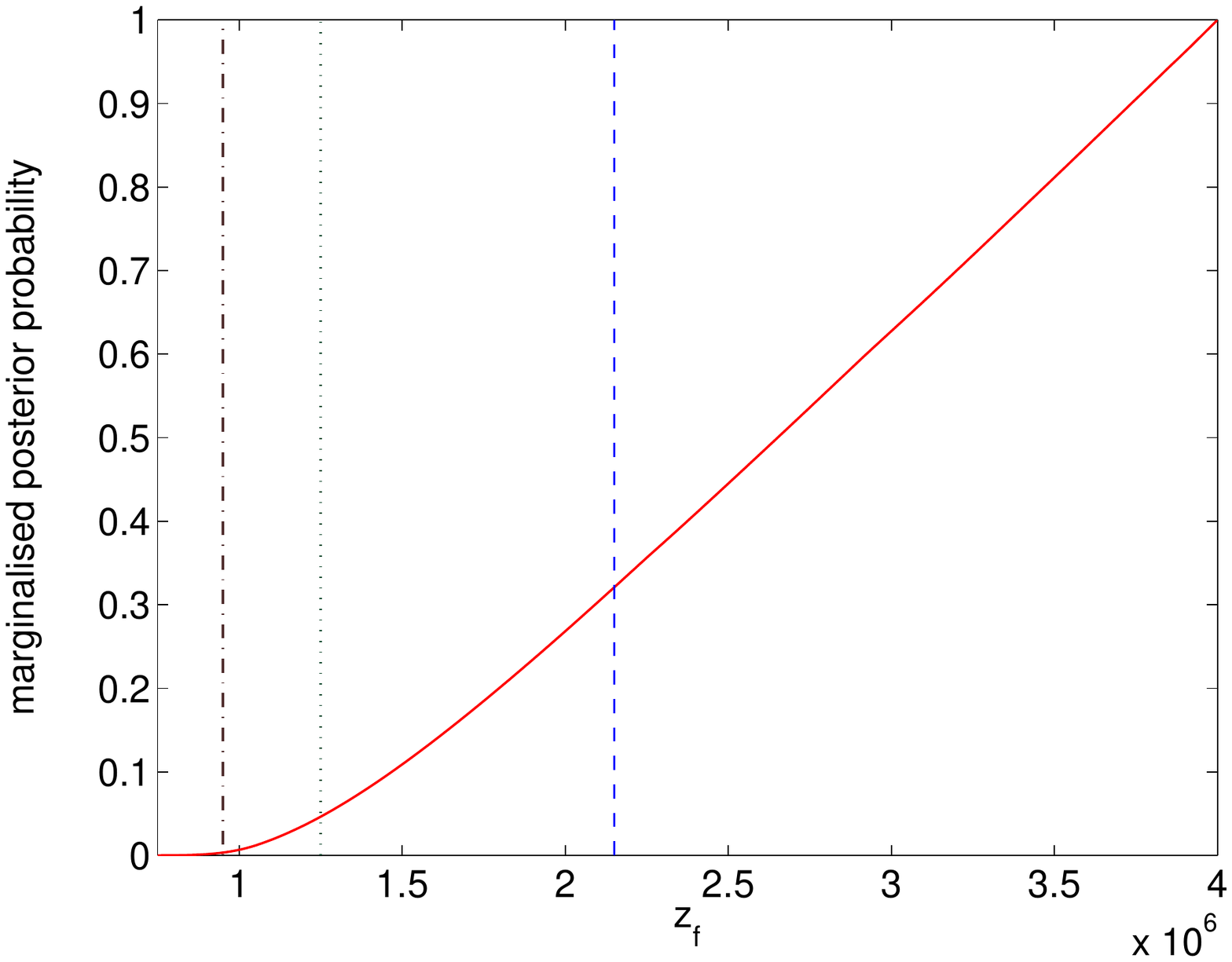}
  \label{fig:3b}
\end{subfigure}
\caption{Marginalized posterior probability for $z_f$ from SDSS data (Left Panel) and  Lyman-$\alpha$ data (Right Panel) are shown. The dashed (blue), 
dotted (green) and dot-dashed (brown) lines 
 indicate the 68$\%$, 95.4$\%$ and 99.73$\%$ regions, respectively.}
\label{fig:3}
\end{figure}

  The marginalized posterior probabilities  for $z_f$  are 
shown in Figure~\ref{fig:3}. We note that the  temperature 
of the universe corresponding to $z = z_f$ from the two data sets
is in the range $30\hbox{--}500 \, \rm eV$. These lower limits 
on the transition temperature  are far below   the constraints
on production redshifts  in the warm dark matter models; in such models a 
dark matter particle with mass $m > 1 \, \rm keV$ is invoked \cite{Viel:2013fqw} and the production redshift lies before the epoch of BBN at a temperature $ T \simeq \rm {MeV} $.
  
\begin{figure}
\begin{subfigure}{.5\textwidth}
  \centering
  \includegraphics[width=1.0 \linewidth]{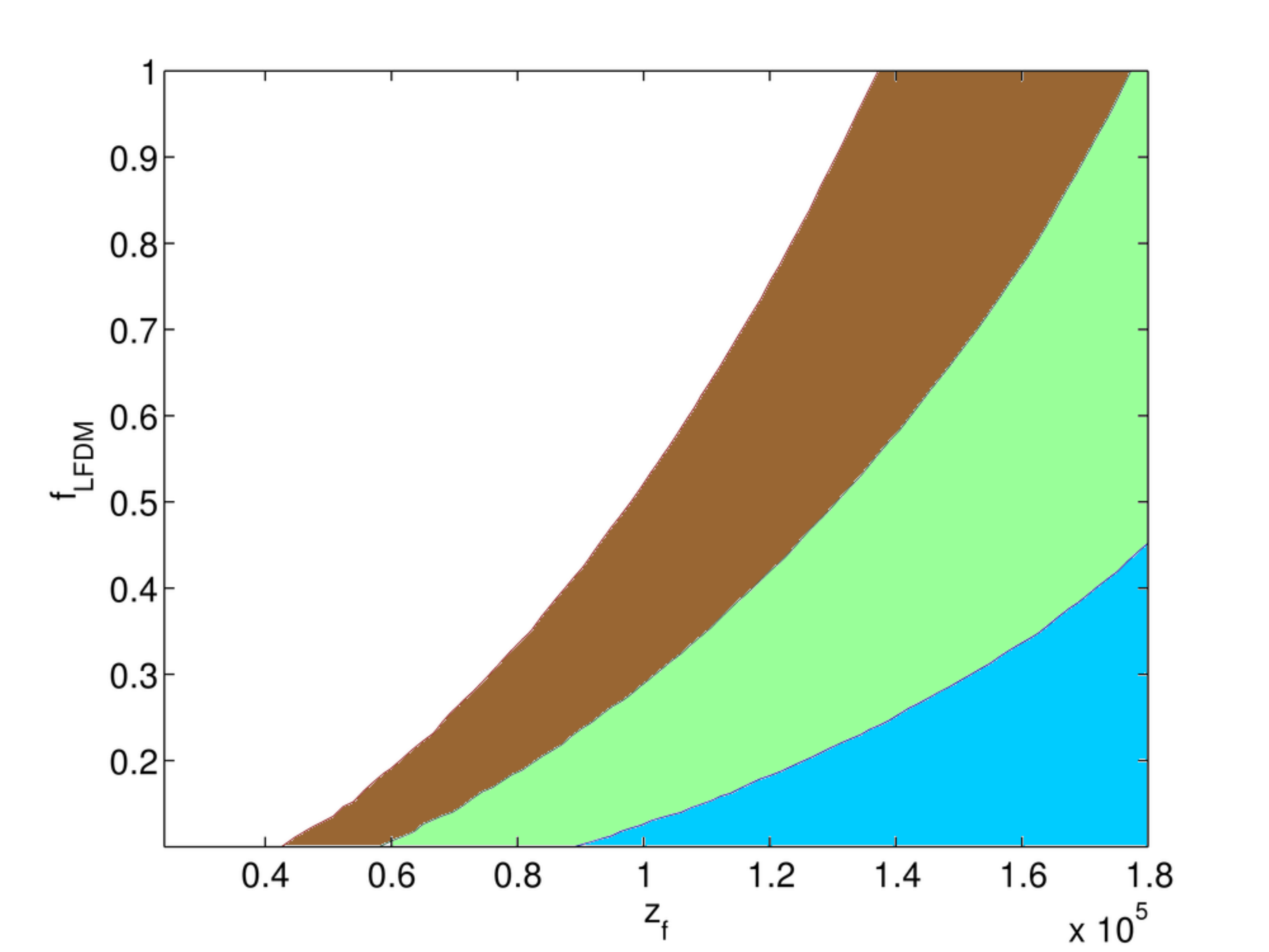}
  \label{fig:4a}
\end{subfigure}%
\begin{subfigure}{.5\textwidth}
  \centering
  \includegraphics[width=1.0 \linewidth]{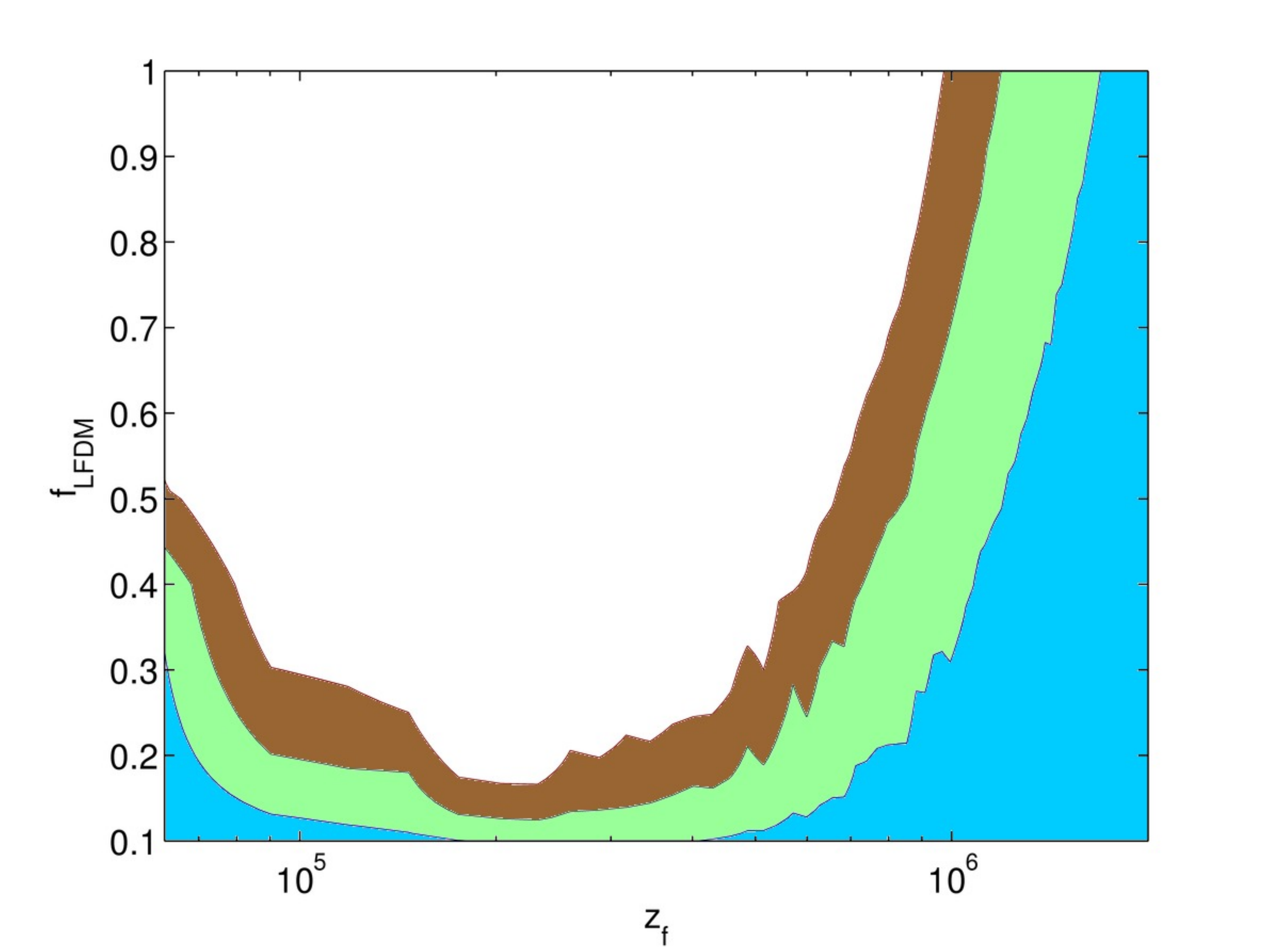}
  \label{fig:4b}
\end{subfigure}
\caption{Contours of $z_f$ and $f_{\rm lfdm}$ obtained using SDSS data  (Left Panel) and the Lyman-$\alpha$ data (Right Panel).}
\label{fig:4}
\end{figure}
\begin{figure}[h!]
\centering
\includegraphics[height=4.5in, width=4.5in]{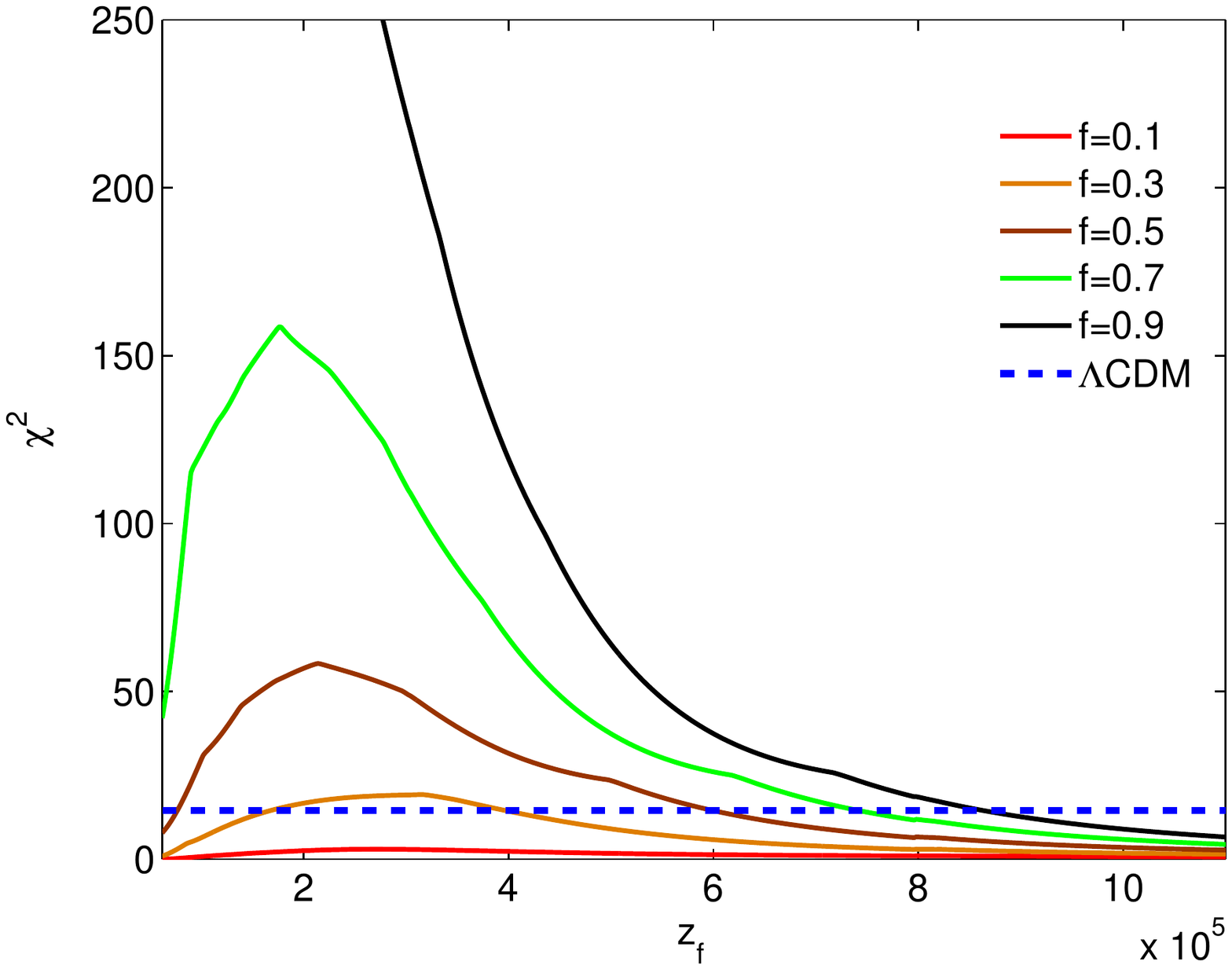}
\caption{$\chi^2$ is plotted as a function of $z_f$ for many different values
of $f_{\rm lfdm}$ for the Lyman-$\alpha$ data. The dotted horizontal 
line is the $\chi^2$ for the $\Lambda$CDM model.}
\label{fig:5}
\end{figure}
\begin{figure}[h!]
\centering
\includegraphics[height=5.5in, width=5in]{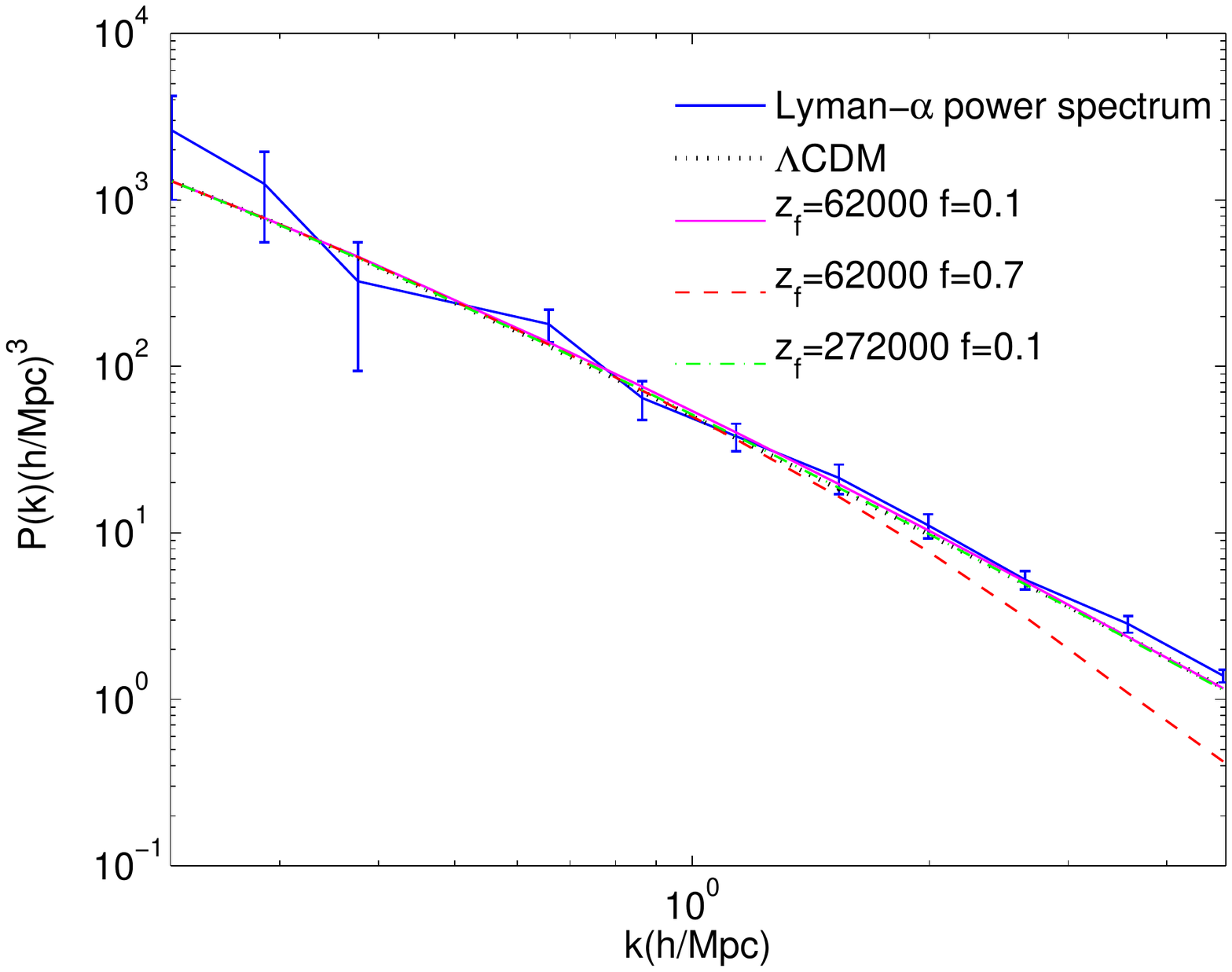}
\caption{The band-powers corresponding to linear matter power spectrum 
extracted  from Lyman-$\alpha$ data are  shown  \cite{Croft:2000hs,Tegmark:2002cy}. Also shown are 
the $\Lambda$CDM model  and three LFDM models for  a range of $z_f$ and 
$f_{\rm lfdm}$.}
\label{fig:6}
\end{figure}

We next consider the case $f_{\rm lfdm} < 1$. Or only a fraction of the 
CDM observed at the present originated at $z = z_f$. This expands the parameter
space under consideration and yields more interesting results. In Figure~\ref{fig:4}, we show  $z_f$--$f_{\rm lfdm}$ contour plots  after marginalizing 
over $N_{\rm eff}$ and the overall normalization $C$.  The SDSS data gives results
similar to the previous case with slightly looser bound on $z_f$. The Lyman-$\alpha$ data, on the other hand, results in very different outcome, as compared to
the earlier case. The $z_f$--$f_{\rm lfdm}$ plain splits into two separate regions in
this case. The region corresponding to $z_f < 10^5$ is ruled out by the SDSS
data but is unconstrained by the Lyman-$\alpha$ data. This underlines the 
importance of using two data sets at different scales for our analysis. Larger
values of $z_f$ is allowed by both the data sets. Further, the Lyman-$\alpha$ 
data results in a better fit as compared to the $\Lambda$CDM case, as seen 
in Figure~\ref{fig:5}, for a large range of values of $f_{\rm lfdm}$ (this 
inference is nearly independent  of $N_{\rm eff}$). 
In particular, $f_{\rm lfdm} = 0.1$ results in a better fit for the entire range
of $z_f$. To understand this improvement of the fit, we show the Lyman-$\alpha$ 
data alongside many theoretical models in Figure~\ref{fig:6}. 
While $\chi^2 \simeq 11$ for the $\Lambda$CDM models, it reduces to 3.5 for 
many models for $f_{\rm lfdm} = 0.1$. This improvement is largely owing to the 
two data points for the largest $k$. This shows the importance of using the 
small scale data for unraveling the nature of LFDM models.

Our analysis clearly shows that LFDM models with a non-zero  $f_{\rm lfdm}$ provide  
a better fit to the data. However, while significant, 
our results need further explanation. In our analysis we assume many 
cosmological parameters to be fixed to their Planck best-fit values. Within
the framework of spatially-flat $\Lambda$CDM model, the relevant 
cosmological parameters---$\Omega_{\rm CDM}h^2$, $\Omega_Bh^2$, $h$, $n_s$---have
been estimated at unprecedented precision \cite{Ade:2013zuv}. For a given angular scale $\ell$, 
the CMBR anisotropies receive dominant contribution from three-dimensional
scales $k$  such that $\ell \simeq  k \eta_0$; $\eta_0 = 13670 \, \rm Mpc$ for the best-fit Planck parameters. As Planck measures  CMBR anisotropies for $\ell < 2000$, the smallest scale to make significant contribution to these observations is $k \simeq 0.15 \, \rm Mpc^{-1}$, which lie in the range of scales
probed by SDSS; Planck results are  compatible with the SDSS DR7 data we use in this  paper (Figure~20 of \cite{Ade:2013zuv}).  However,   these scales 
are  larger than the scales involved in
Lyman-$\alpha$ measurements and therefore the   Lyman-$\alpha$ data gives us 
independent information of the matter  power spectrum on scales not probed by
Planck. This also means that we are justified in assuming priors on cosmological parameters from Planck, even though we still need to explore
the whole range of parameters allowed by Planck to put our result on a firmer
footing. 

\begin{figure}[h!]
\centering
\includegraphics[height=4.5in, width=4.5in]{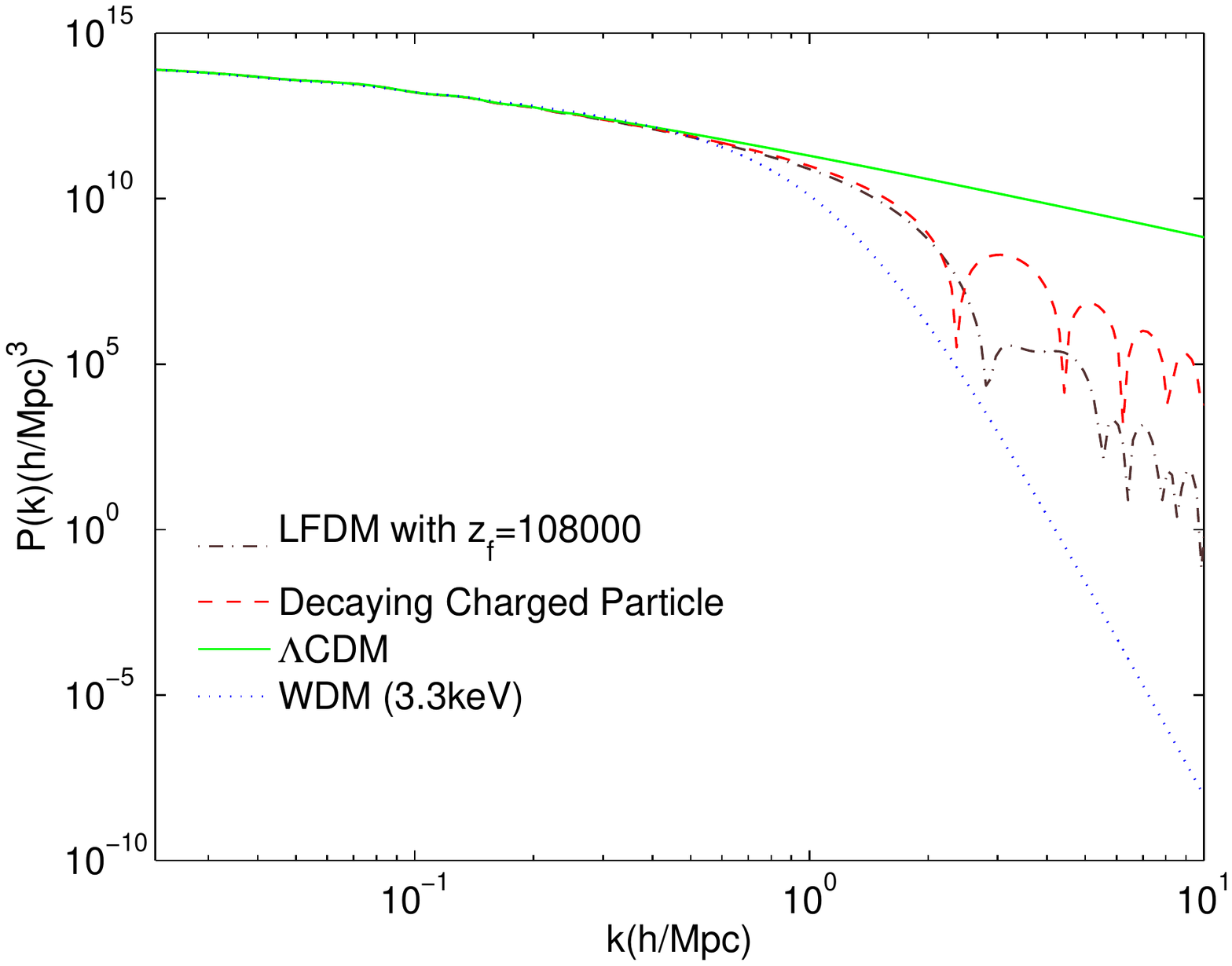}
\caption{The LFDM results  are  compared to  other models  that also predict 
 suppression of  matter power at small scales. The power spectra  (unnormalized)
for  WDM and the charged 
decay  models follow from \cite{Viel:2013fqw} and \cite{Sigurdson:2003vy}, 
respectively. In the charged decay model, the decay  redshift is the 
same as the formation redshift for  the LFDM model shown in the Figure.}
\label{fig:7}
\end{figure}

In Figure~\ref{fig:7} we compare the predictions of our model  with other models
in which the power is suppressed at small scale with respect to the $\Lambda$CDM model. One such model is the WDM model; in Figure~\ref{fig:7} we show the matter
 power spectrum for a much-studied WDM model \cite{Viel:2013fqw}. In WDM 
models, the neutrino-like particle is much heavier than the usual standard model
neutrino and  its mass density is matched to the present day dark matter
mass density. In such class of models, the WDM component can begin to cluster
after it become non-relativistic at a time when  $T \lesssim m_{\rm WDM}$. However, owing to the fact that the WDM particle
remains semi-relativistic for a long time after this era, the matter power 
is suppressed on a large range of scales that enter the horizon during 
this period (for details see e.g. \cite{2013neco.book.....L}). In this case, we get suppression without oscillations for scales of interest.

The other model of interest is the decay of a charged particle into a CDM
particle after BBN \cite{Sigurdson:2003vy}. In this model, the initial conditions for the CDM perturbations  are derived  from the tightly coupled photon-Baryon Plasma. We compute the impact of 
of this process  on the resultant power spectrum  by assuming  the decay to be a sharp transition. We compare the 
matter power spectrum in this case with the LFDM model in Figure~\ref{fig:7} for the same 
formation  redshift in both cases. As noted above, the initial conditions for 
the LFDM model come from  massless neutrinos (Figure~\ref{fig:1a}). The 
impact of 
the  difference of
 initial conditions in the two cases   can clearly be seen in Figure~\ref{fig:7}: for LFDM
model  neutrino oscillations follow an exponential slope below the scale 
of the horizon entry at the formation redshift. However, for the charged decay
particle,  the photon-baryon plasma oscillate  inside the horizon with 
characteristic scales determined  by the sound velocity of the coupled 
photon-baryon fluid and its decay scale is determined by Silk damping
(for details see Figure~3 of \cite{Ma:1995ey} and the discussion preceding it).

 \section{Discussion}
  
    In this work we have investigated  the epoch of dark matter formation in the universe  for a class of non-standard (non-WIMP)   dark matter scenarios. Especially we have studied how late the dark matter can form. Unlike the case of electroweak  WIMP where dark matter formation happens through thermal freeze-out at a temp 
    $T \simeq  \rm {GeV}$,  in our models   the dark matter formation happens considerably after BBN  but  before the   CMBR decoupling. Our study  is mainly inspired by  a few viable models of "late forming dark matter"  \cite{Das:2006ht, Sigurdson:2003vy}. In such models the matter power is suppressed at small
scales which can be probed by cosmological observables at low redshifts
using the available data on the linear power spectrum.

In the present study  we confront models of  LFDM with the existing SDSS  data
on galaxy power spectrum and the linear power spectrum extracted from Lyman-$\alpha$ data for $z > 2$. Our results can be summarized as follows: (a) 
if all the presently observed CDM  is late forming then both the data sets result
in upper limits on the redshift of formation of LFDM, with Lyman-$\alpha$ data
resulting in tighter bounds: $z_f < 3 \times 10^6$ (99\% confidence limit) (Figure~\ref{fig:2}), (b)
if we allow only a fraction of the CDM to form at late times, then we improve
the quality of fit as compared to the $\Lambda$CDM model 
 for the Lyman-$\alpha$ data. This is suggestive that the present data 
allows for a fraction of the CDM to form at $z_f \simeq 10^5$ (Figure~\ref{fig:4}). In particular
our result underlines the importance of the Lyman-$\alpha$ data  for our study. 
In the recent past, the quantity of Lyman-$\alpha$ data available has sharply increased with the ongoing survey SDSS-III/BOSS \cite{Palanque-Delabrouille:2013gaa}; 
and the results from this survey are expected to throw further light on the 
models of LFDM.  We hope to return to this issue with as the new data becomes available.  

  We compare the predictions of our model with a few 
 well-studied models that also result in a suppression of matter 
power at small scales (Figure~\ref{fig:7}). Different models---LFDM, 
charged decay model, and WDM---result in varying scales of characteristic 
oscillations and decay. The upcoming data might enable one to distinguish 
between these different deviations from the standard $\Lambda$CDM model
and we hope to return to this issue in the near future. 
  
Another possible  future direction involves doing  detailed N-body simulation of this model  to see if there is any specific signatures of LFDM 
at  small scale non-linear structure formation.  Another interesting 
 study could be   the effect of LFDM on the epoch of  the  formation of 
first stars. Both of these studies are  beyond the scope of this paper and have been kept for the  future work.
 
 \section{Acknowledgment}
We would like to thank  Anze Slosar for his useful suggestions and comments and also to  Daniel  Boriero for the help with
the SDSS power spectrum data. We thank Pier Stefano Corasaniti for sending useful comments.

\bibliography{mybib}{}
\bibliographystyle{JHEP}

\end{document}